\documentclass[a4paper]{article}

\usepackage{icrc2013}

\title{Simulation study on origin of multi-core events in cosmic rays extensive air showers}

\shorttitle{ICRC 2013}

\authors{
Yonggang Luo$^{1,2}$,
Shuwang Cui$^{3}$,
Xinhua Ma$^{4}$,
Jing Zhao$^{4}$,
Cunfeng Feng$^{2}$
}

\afiliations{
$^1$ Taishan College, Shandong University, Jinan, Shandong, China \\
$^2$ Department of Physics, Shandong University, Jinan, Shandong, China\\
$^3$ The college of Physics Science and information Engineering, Hebei Normal University, Shijiazhuang, Heibei, China\\
$^4$ Institute of High Energy Physics, Chinese Academy of Science, Beijing, China \\
}

\email{luoyg@mail.sdu.edu.cn}

\abstract{
Some experiments have found multi-core events in cosmic rays extensive air showers which should be interpreted by hadronic interaction theory. In this paper, the multi-core events are reproduced by Monte Carlo simulation with CORSIKA. The origin of each sub-cores is tracked back from the observation level. The interaction mechanism and original particles of sub-core are studied in this paper.
}


\begin{document}
\maketitle

%Begin a section.
\section{INTRODUCTION}

Multi-core events is still a puzzle in our current understanding of the cosmic rays extensive air shower. The results from some experiments (Mt. Norikura, Mountain Emulsion Chambers, and hardonic calorimeter EASTOP as some examples) point out that the multi-core events can be studied by ground-based observatory.\cite{bib:Munakata,bib:Hazen,bib:Matano,bib:Bakich,bib:Bosia,bib:Chudakov,bib:Ren,bib:Lattes,bib:Coll,bib:Aglietta} At present, the ARGO-YBJ experiment \cite{bib:ARGO} gives a unique observation on the multi-core showers\cite{bib:mcore} with large distance between the main core and sub-core thanks to its full coverage RPCs detector. The traditional view for this phenomenon is jet production, which is essentially produced in the process of the leading particle interacts with "air" target nuclei. Therefore we can get some information about hadronic interaction models from multi-core events.

Some Mountain Emulsion Chambers experiments show that the high values for the physical parameter ${\chi}=\sqrt{E_1E_2}r_{12} > 1000$ TeV cm, where $E_1$, $E_2$ are the energies of two cores and $r_{12}$ is distance between them. Up till now, it is difficult to apply hadronic interaction models to explain this phenomenon.\cite{bib:Czhen,bib:CZhen} At the same time, this high $\chi$ value events cannot be deviated certainly from Monte Carlo simulations.\cite{bib:Hwang} The events with high $\chi$ value at large distance cannot be studied well on the Mountain Emulsion Chambers experiments due to limited active area, but the ARGO-YBJ experiment with large active area creates a new platform for studying the multi-core events and some hadronic interaction models. In this work, the multi-core events were studied by Monte Carlo simulation.

\section{EXTENSIVE AIR SHOWERS SIMULATION}

In this simulation, cosmic rays are generated by CORSIKA\cite{bib:Corsika} ({\bf C}osmic {\bf R}ay {\bf SI}mulations for {\bf KA}scade) version CORSIKA-6981. The selected hadronic interaction model is QGSJETII-GHEISHA. QGSJET model describes high-energy hadronic interaction using the quasi-eikonal Pomeron parameterization for the elastic hadron-nucleon scattering amplitude. GHEISHA model describes hadronic collisions up to some 100 GeV. To get more information, the other used options are SLANT option, EHISTORY option and ROOTOUT option. The SLANT option gives the longitudinal distribution in slant depth bins along the shower axis. The EHISTORY option gives the addition information on the prehistory of muons (i.e. Some information of mother particles and grandmother particles of this muon). The ROOTOUT option transmits the particle output to an output "DATnnnnnn.root" file in root format and gives some convenience for analyzing data.

Primary energy is set from 100TeV to 10PeV. The composition of primary nuclei consists of protons, He, MgAlSi, CNO and Fe. Primary energy spectrum used the Horadel model \cite{bib:Horandel}. Zenith angle is $0^{\circ}-45^{\circ}$, and azimuth angle is $0^{\circ}-360^{\circ}$. Observation level is at YangBaJing (4300m above sea level).

\begin{figure}[t]
  \centering
  \includegraphics[width=0.4\textwidth]{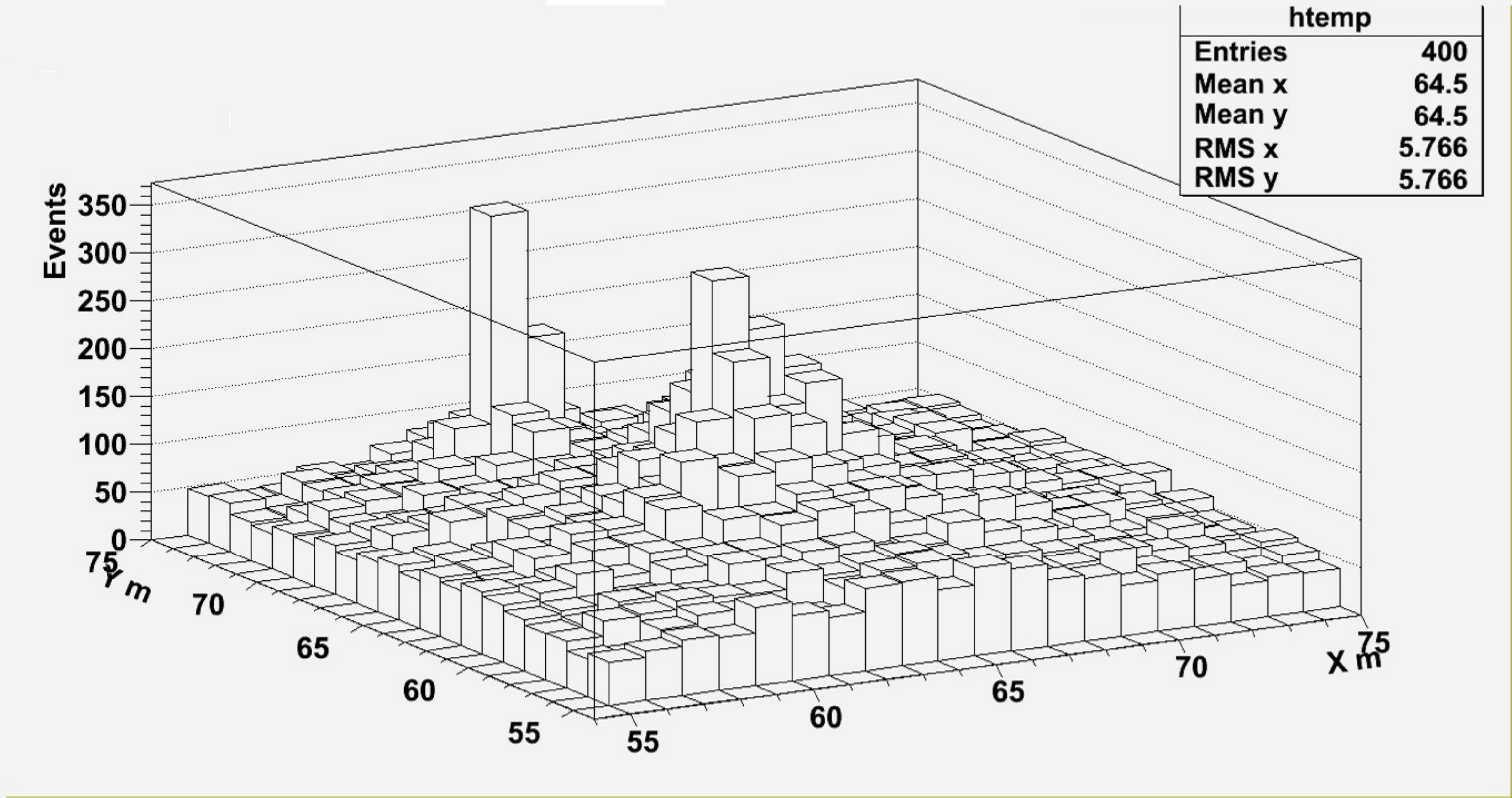}
  \caption{A multi-core event produced from CORSIKA.}
  \label{simp1}
 \end{figure}

For each multi-core event (Fig.\ref{simp1} is an example) produced from CORSIKA, one CORSIKA tool named as corsikaread\underline{ }history\cite{bib:corsikaread} is used to search the grandmother particles and mother particles of muons which found at observation level. This tool is developed to read on the addition information of grandmother particles and mother particles of muon particles.

In order to observe the development of secondary core, the program COAST\cite{bib:Coast} ({\bf CO}rsika d{\bf A}ta acces{\bf S} {\bf T}ools is a bundle of C++ code providing simple and standardized access to CORSIKA data ) version COAST-V4R3 is applied to find multi-core events produced from our simulation and their longitudinal development.

\section{ANALYSIS AND RESULTS}

\subsection{MUONS INSIDE THE SUB-CORE}
\begin{figure}[t]
  \centering
  \includegraphics[width=0.4\textwidth]{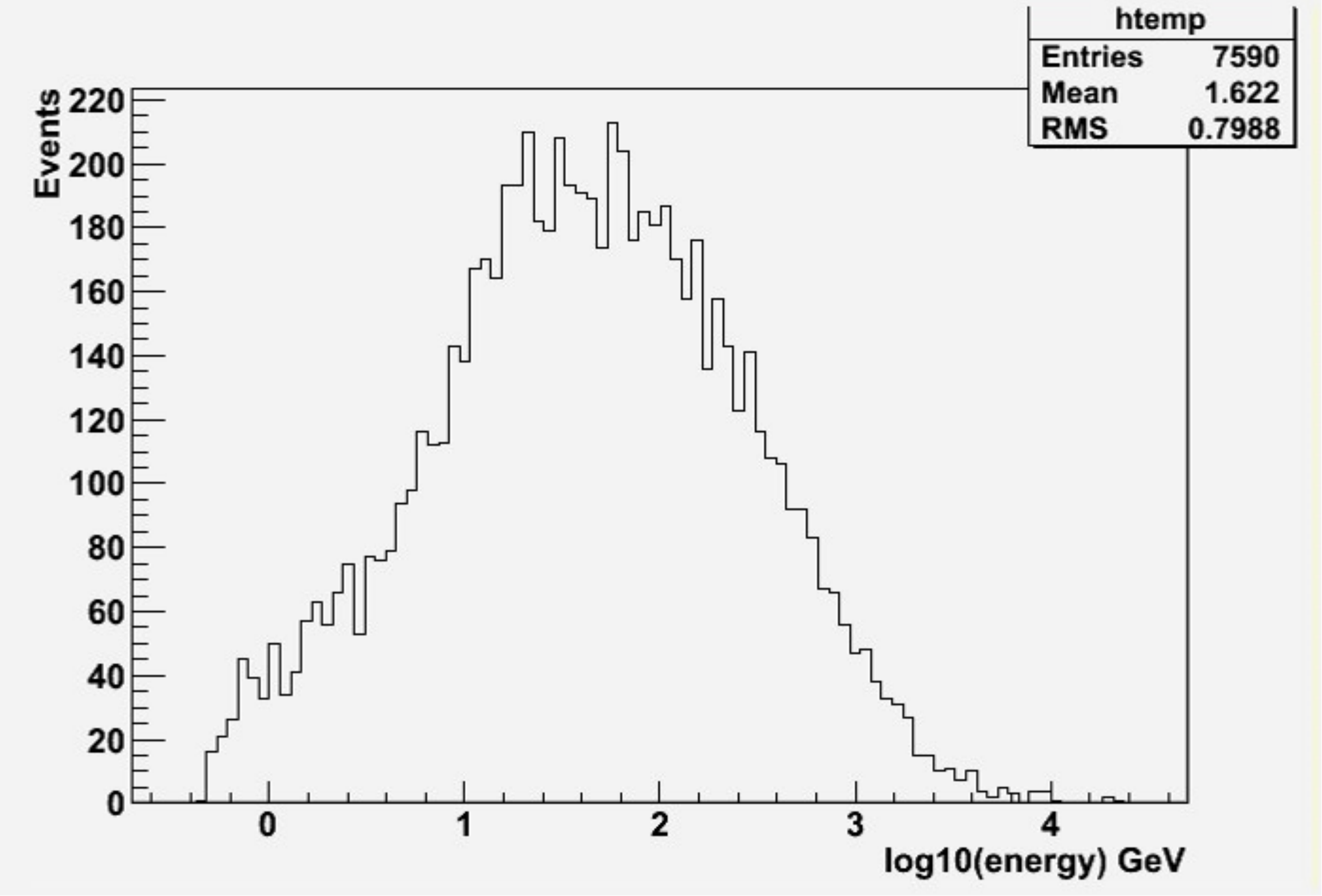}
  \caption{Energy distribution of mother particles of all muons at the observation level in total 25 events with primary energy from 1 PeV to 10 PeV.}
  \label{simp3}
 \end{figure}

\begin{figure}[t]
  \centering
  \includegraphics[width=0.4\textwidth]{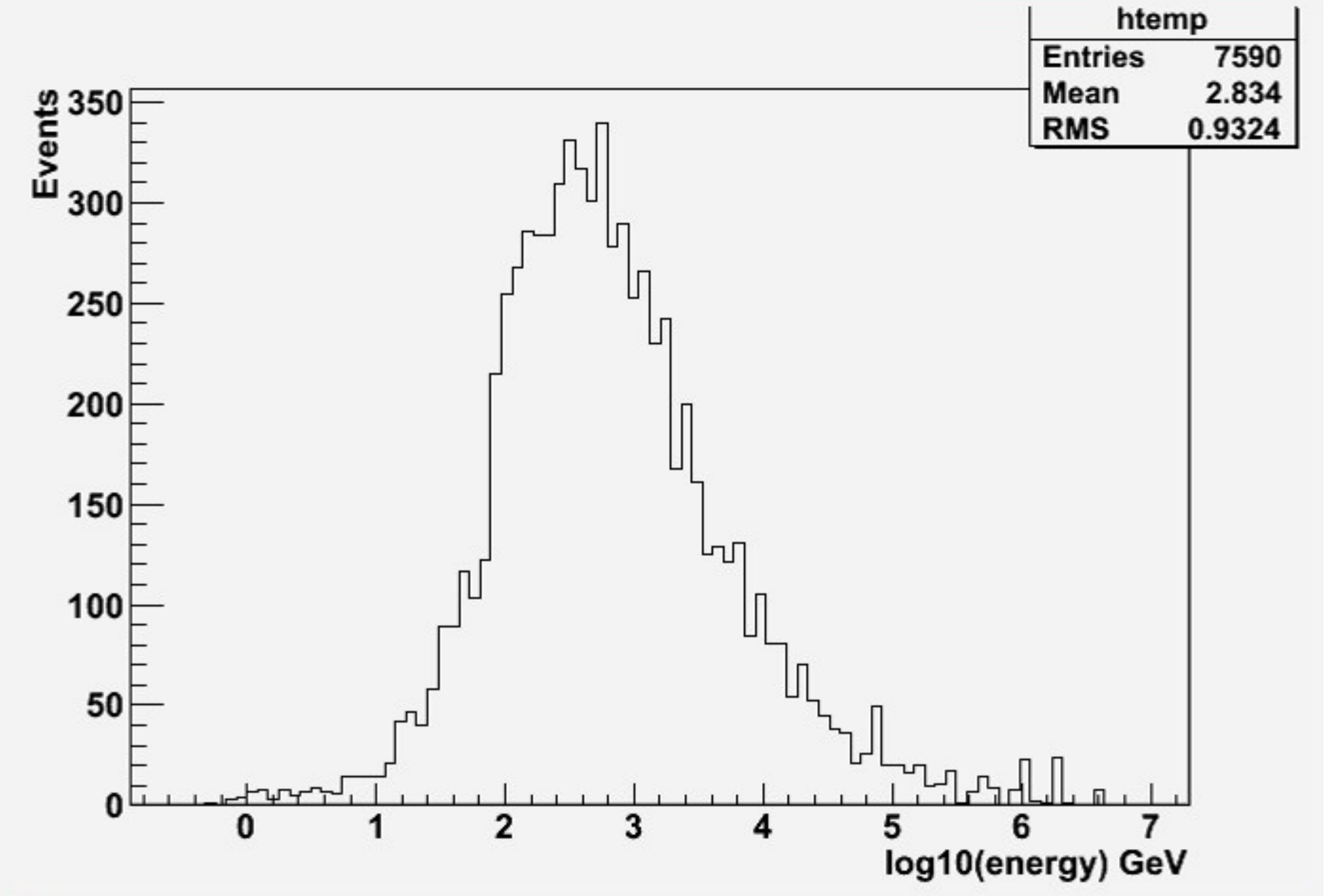}
  \caption{Energy distribution of grandmother particles of all muons at the observation level in total 25 events with primary energy from 1 PeV to 10 PeV.}
  \label{simp4}
 \end{figure}

So far, only 44 multi-core events are produced by CORSIKA due to the multi-core events are rare events, and among them, 30 events have muons in the region of sub-core. For these 30 events, muons are found in the region of sub-core, we read information of the mother particles and grandmother particles of all muons at the observation level. From these information, for example, energy distribution of grandmother particles and mother particles of all muons at the observation level are obtained. (Fig.\ref{simp3} and Fig.\ref{simp4}). Of course, the information for all muons arriving in the region of sub-core can be read out.

\begin{figure}[t]
  \centering
  \includegraphics[width=0.4\textwidth]{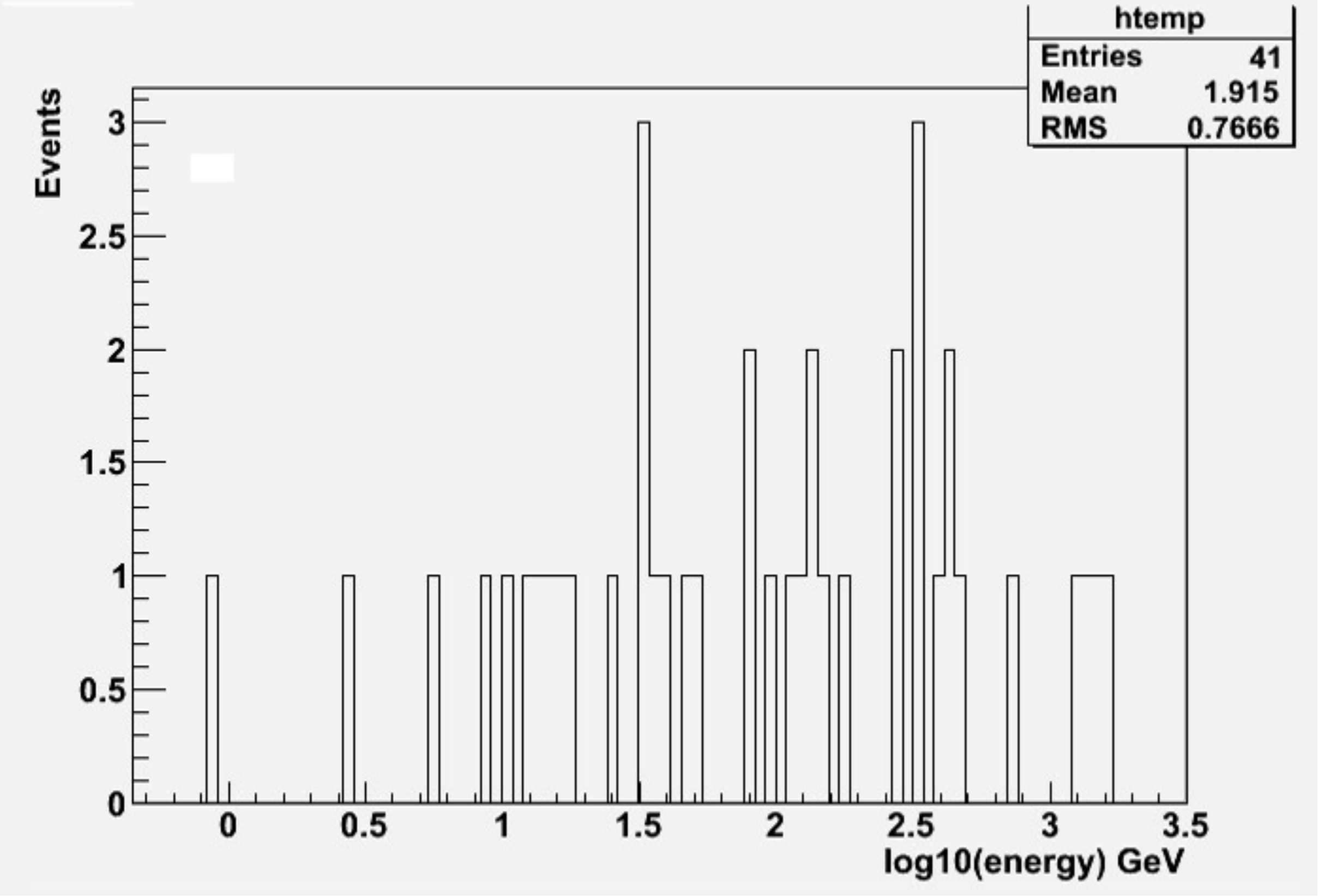}
  \caption{Energy distribution of mother particles of all muons arriving in the region of sub-cores in 30 events with primary energy from 1 PeV to 10 PeV.}
  \label{simp5}
 \end{figure}

\begin{figure}[t]
  \centering
  \includegraphics[width=0.4\textwidth]{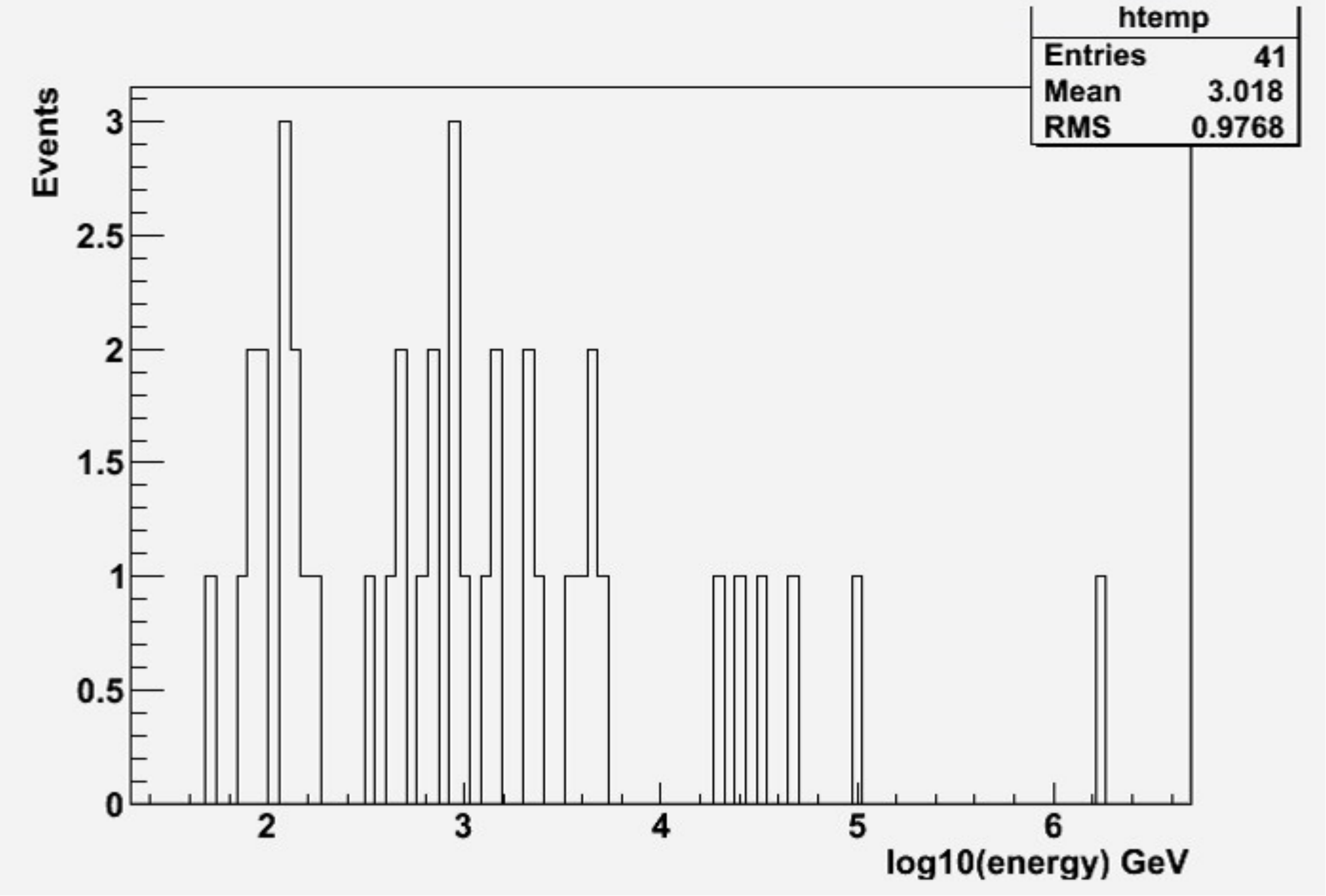}
  \caption{Energy distribution of grandmother particles of all muons arriving in the region of sub-cores in 30 events with primary energy 1 PeV to 10 PeV.}
  \label{simp6}
 \end{figure}

All the information about energy distribution of grandmother particles and mother particles of all muons arriving in the region of sub-core is collected. And then, the same method is applied to produce figure (Fig.\ref{simp5} and Fig.\ref{simp6}) reflecting the energy distribution of grandmother particles and mother particles, where 41 muons was collected in 30 multi-core events.

For the mean value of energy, mother particles and grandmother particles in the sub-core are respectively $10^{1.9}$ GeV and $10^{3.0}$ GeV, while the average energy of mother particles and grandmother particles of all muon (inside/outside sub-core) in 25 events are respectively $10^{1.6}$ GeV and $10^{2.8}$ GeV. The comparison of the former (Fig.\ref{simp3} and Fig.\ref{simp4}) and the latter (Fig.\ref{simp5} and Fig.\ref{simp6}) indicates that the secondary core corresponds higher energy mother particles and grandmother particles. The average value of energy of grandmother particles and mother particles of muons inside the sub-core are higher, but not significantly higher than the average value of all muons in the showers.

\subsection{LONGITUDINAL DEVELOPMENT ANALYSIS AT THE SUB-CORES}

By considering CORSIKA and COAST in our simulation, the extensive air showers development is record at different observation levels. And the EHISTORY option is activated so that the history information of muons also is recorded. As an example: the primary cosmic ray is Nitrogen nucleus with energy 1.3 PeV, zenith angle is $27.8^{\circ}$ and azimuth angle is $83.4^{\circ}$. The sub-core at the observation level has one $\mu^-$. Because EHISTORY option was activated, the mother particles and grandmother particles of this $\mu^-$ were found. The mother particle of the $\mu^-$ is $\pi^-$ and appears at 29770m above sea level with energy 1.5 TeV and disappears at 28156m above sea level. The grandmother particle of the $\mu^-$ is neutron and appears at 35508m above sea level with energy 93.92 TeV and then interacts with air nuclei to some secondary particles at 29770m above sea level. And then we can get the interaction process from generation counter as follow:

Nitrogen + Air $\rightarrow$ Neutron + recoil nucleus,

Neutron + Air $\rightarrow$ $\pi^-$ + recoil nucleus,

$\pi^-$ $\rightarrow$ $\mu^-$ + $\bar{\nu}$.

\begin{figure}[t]
  \centering
  \includegraphics[width=0.4\textwidth]{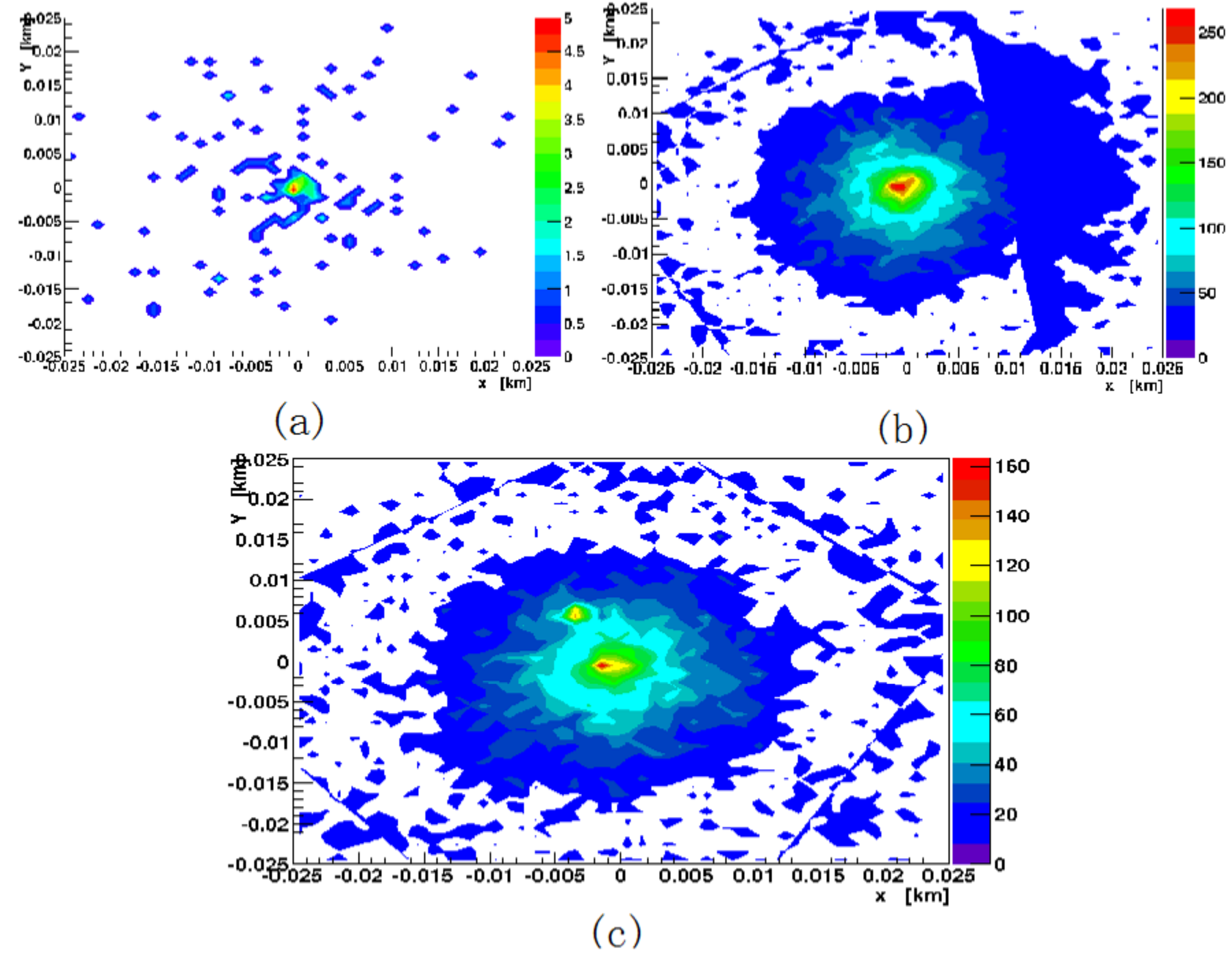}
  \caption{(a) Distribution of electrons at disappear level (28516m above sea level) for mother particles. (b) Distribution of electrons at appear level (about 5200m above sea level) for sub-core. (c) Distribution of electrons at YangBaJing level (4300m above sea level). }
  \label{simp2}
 \end{figure}

With COAST help, the longitudinal development of extensive air showers is recorded at different observation levels. Fig.\ref{simp2} is an example of extensive air showers at 3 levels. The distribution of sll type of particles at any height above observation level (Fig.\ref{simp2}) can be found. These figures provide us an opportunity to observe clearly the development of multi-core events.

In this work, we focus on the distribution of electron and positron at some levels between the birth point of grandmother particles and the disappear point of mother particles. We have analyzed three events including the event mentioned above up to now. As a result, we find that there is no sub-core at the level of mother particles decay to muons.

As a discussion, we notice that the mother particle pion have much high energy (1.5 TeV). Thus we surmise that the pion emits a high energy gamma before it decays to a muon at 28156m above sea level, and then the sub-core is generated by the high energy gamma at the lower altitude (about 5200m above sea level). Analysis on more events will give a more certain conclusion in the future.

\section{SUMMARY AND CONCLUSION}

A Monte Carlo simulation has produced some multi-core events and then we find the character of multi-core events by reading the information of grandmother particles and mother particles. The origin of secondary core are tracked back from the observation level, and then we observe clearly the change of sub-core. So far we have no evidence to confirm convention hadronic interaction models, but researching on more simulation events will provide details about the hadronic interaction models in the future.

\vspace*{0.5cm}
\footnotesize{{\bf Acknowledgment:}{This work is supported by Funding from the MOE, China, and by Funding for Fostering Talents in Physics Science Foundation of China J1103212.}}

\end{document}